\documentclass{ws-p8-50x6-00}

\usepackage{graphicx}
\usepackage{psfig}
\usepackage{indentfirst}

\newcommand{\p}{$\phi$}
\newcommand{\slope}{$T$}

\newcommand{\ct}{379$\pm$51}
\newcommand{\mt}{369$\pm$73}
\newcommand{\lt}{417$\pm$76}
\newcommand{\cy}{5.73$\pm$0.37}
\newcommand{\my}{3.33$\pm$0.38}
\newcommand{\ly}{0.98$\pm$0.12}

\newcommand{\cphipi}{0.021$\pm$0.001}
\newcommand{\mphipi}{0.019$\pm$0.002}
\newcommand{\lphipi}{0.019$\pm$0.002}

\newcommand{\ctsu}{$\pm$ 45}

\newcommand{\cysu}{$\pm$ 0.57}

\newcommand{\cphipisu}{$\pm$ 0.004}

\begin{document}

\title {Resonance Production at STAR}

\author{Eugene T. Yamamoto}

\address{Lawrence Berkeley National Laboratory, Berkeley, CA 94720, USA\\
E-mail: ETYamamoto@lbl.gov}


\maketitle

\abstracts{
We present the first measurement of mid-rapidity $\phi$ vector meson production in
$Au+Au$ collisions at RHIC ($\sqrt{s_{_{NN}}} =130$~GeV) from the STAR detector.
For the 11\% most central collisions, the slope parameter from an exponential fit to
the transverse mass distribution is \slope$=$~\ct(stat)\ctsu(syst)~MeV,
the yield $dN/dy$ $=$ \cy(stat)\cysu(syst) per event and the ratio $N_{\phi}/N_{h^-}$
is found to be
\cphipi(stat)\cphipisu(syst). We currently place the value of the $N_{\phi}/N_{K^-}$ 
ratio between 0.10 and 0.16.
The measured ratios $N_{\phi}/N_{h^-}$ and $N_{\phi}/N_{K^-}$,
as well as $T$ for the $\phi$ meson at mid-rapidity do not change for the selected centrality
bins.}

\section{Introduction}
The central topic of relativistic heavy ion physics is the study of
Quantum Chromodynamics (QCD) in extreme conditions of high temperature and 
high energy density over large volumes~\cite{wilczek}.
Vector mesons may probe the dynamics of particles and chiral symmetry~\cite{chiral}
in relativistic
heavy ion collisions: their production mechanisms and subsequent dynamical 
evolution have been a topic of experimental investigation
~\cite{na49phi,na50phi,e802,e917qm99}. The $\phi$ meson is of particular interest
due to its $s\bar{s}$ valence quark content, which may make the $\phi$
sensitive to strangeness production from a possible early partonic phase~\cite{rafelski,shor,koch}.

In central $Pb+Pb$ collisions at the CERN SPS (nucleon-nucleon center of mass energy
$\sqrt{s_{_{NN}}} \simeq 17$~GeV), the slope
parameter ($T$) in an exponential fit to the transverse mass ($m_t$)
distribution at mid-rapidity ($\propto e^{-m_{t}/T}$) follows a systematic trend as a function
of hadron mass for pions, kaons and protons~\cite{na44}. 
This observation is indicative of a common expansion velocity developed in the final
state for pions, kaons, and protons~\cite{uheinzflow}.
The slope parameters, however, 
measured for multi-strange hyperons $\Xi$ and $\Omega$
~\cite{wa97}, and for J/$\psi$~\cite{na50qm01} show deviations from a linear mass dependence, 
suggesting that these particles do
not interact as strongly in the final state at SPS energies~\cite{RQMD}.
Measurements of $\phi$ meson production at the SPS were 
inconclusive~\cite{na49phi,na50phi}. 
Significantly different values for the $\phi$ slope parameter have been obtained from exponential
fits to the measured $m_t$ spectra in central $Pb+Pb$ collisions when using the
$K^+K^-$ decay channel \cite{na49phi} and when using the $\mu^+\mu^-$ decay channel
\cite{na50phi} of the $\phi$ meson.
This difference, however, is not apparent
in peripheral collisions \cite{friese01,quintans}. Possible scenarios to explain the difference
have been discussed in the literature \cite{johnson,soffqm00}.

We report the first measurement of mid-rapidity ($|y| < 0.5$)
$\phi$ production in $Au+Au$ collisions at RHIC ($\sqrt{s_{_{NN}}}=130$~GeV) via
the $\phi \rightarrow K^+K^-$ decay channel (branching ratio $= 0.491$)
using the Solenoidal Tracker At RHIC (STAR) detector \cite{star}.
Systematics of $\phi$ meson  production as a function of centrality at RHIC as well
as its $\sqrt{s_{_{NN}}}$ dependence will be discussed. 
\section{Analysis}
The STAR detector consists of several
detector sub-systems in a large solenoidal analyzing magnet.  For the
data taken in the year 2000 and presented here, the experimental 
setup consisted of
a Time Projection Chamber (TPC), a Central Trigger Barrel
(CTB), and two Zero Degree Calorimeters (ZDC) located upstream 
along the beam axis.  The TPC is a cylindrical drift
chamber with multi-wire proportional chamber readout.
With its axis
aligned along the beam direction, the TPC provided complete azimuthal
coverage. Surrounding the TPC was the CTB,
which measured energy deposition from
charged particles. 
The ZDC's measured beam-like neutrons from the fragmentation of
colliding nuclei. The CTB was used in conjunction with the ZDC's as
the experimental trigger.

Data used in this analysis were taken with two different trigger conditions: 
a minimum-bias trigger
requiring a coincidence between both ZDC's and a central trigger
additionally requiring a high hit multiplicity in the CTB. 
The central trigger corresponded to approximately the top 15\% of the 
measured cross section for $Au+Au$ collisions. Data from both the
minimum-bias trigger and central trigger were used for this analysis.

Reconstruction of the $\phi$ was accomplished by calculating 
the invariant mass ($m_{inv}$), transverse momentum ($p_t$), 
and rapidity ($y$) of all permutations 
of candidate $K^+K^-$ pairs. The resulting $m_{inv}$ distribution 
consisted of the $\phi$ signal as well as combinatorial background. 
The shape of the combinatorial background was calculated using
the mixed-event technique \cite{mixing1,mixing2}. 

For the centrality measurement, the raw total charged multiplicity
distribution within a pseudo-rapidity window
$| \eta | \le$ 0.75 was divided into three bins corresponding to 
85--26\%, 26--11\% and the top 11\% of the measured cross section for
$Au+Au$ collisions \cite{starflow1,crosssection,harrisqm01}.
Events were selected with a primary vertex z position ($z$) from
the center of the TPC of $|z| < 80$~cm.
These events were further divided according to $z$ in 16 bins,
and event mixing was performed for events within
each bin to construct background distributions with reduced acceptance-induced
distortions in the mixed-event background.
Consistent results were obtained
when we constructed the background distribution using like-sign pairs from the same
event.
\begin{figure}
\centering\mbox{
\psfig{figure=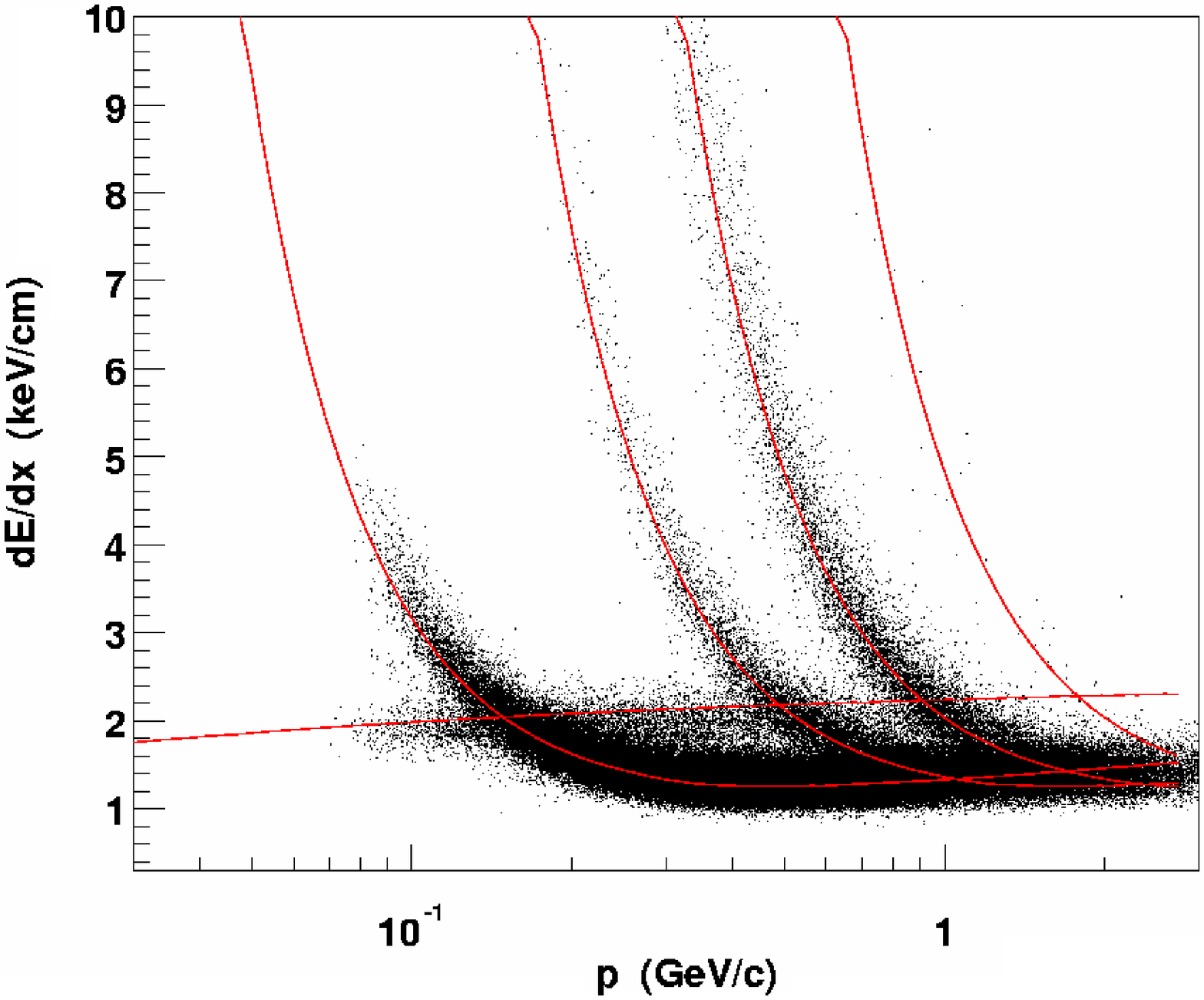,width=.6\textwidth}
\psfig{figure=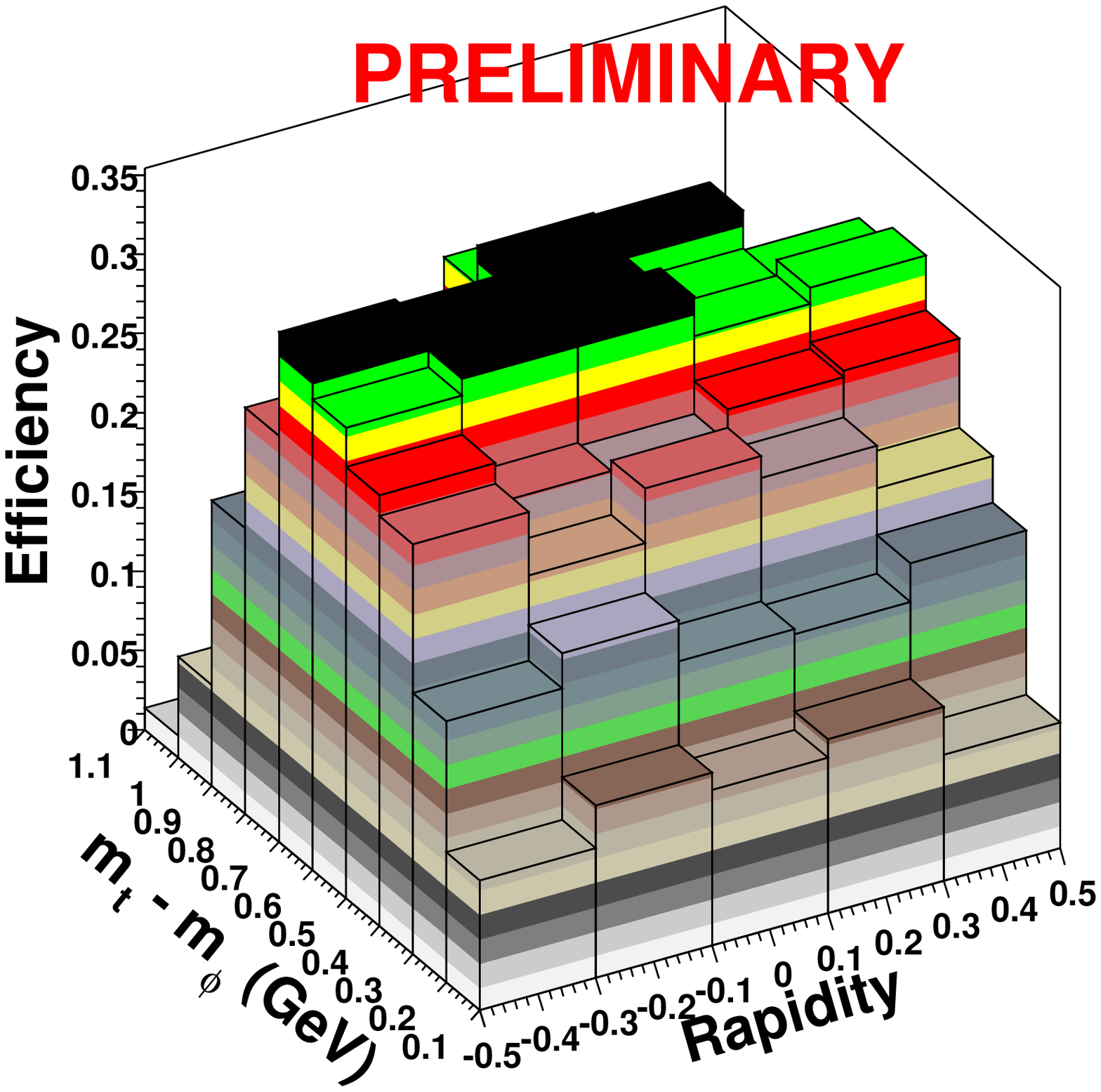,width=.49\textwidth}}
\caption{
\textbf{Left:} The measured $\langle dE/dx \ra$ vs. $p$ for reconstructed tracks in the TPC.
\textbf{Right:} Reconstruction efficiency vs. $m_t - m_{\phi}$ vs. rapidity for minimum-bias events.
}
\label{de_effic}
\end{figure}

Particle identification (PID) was achieved by correlating 
the ionization energy loss ($dE/dx$) of charged particles in the TPC gas
with their measured momentum (left panel of Figure \ref{de_effic}). 
By truncating the largest 30\% $dE/dx$ values along the track, a sample was selected to
calculate the mean $\langle dE/dx \rangle$. 
For the most central events, the average $\langle dE/dx \rangle$ 
resolution was found to be about 11\%.
The measured $\langle dE/dx \rangle$ is reasonably described by the
Bethe-Bloch function smeared with a resolution of width $\sigma$. 
Tracks within $2\sigma$ of the kaon Bethe-Bloch curve were 
selected for this analysis.

\begin{figure}
\centering\mbox{
\psfig{figure=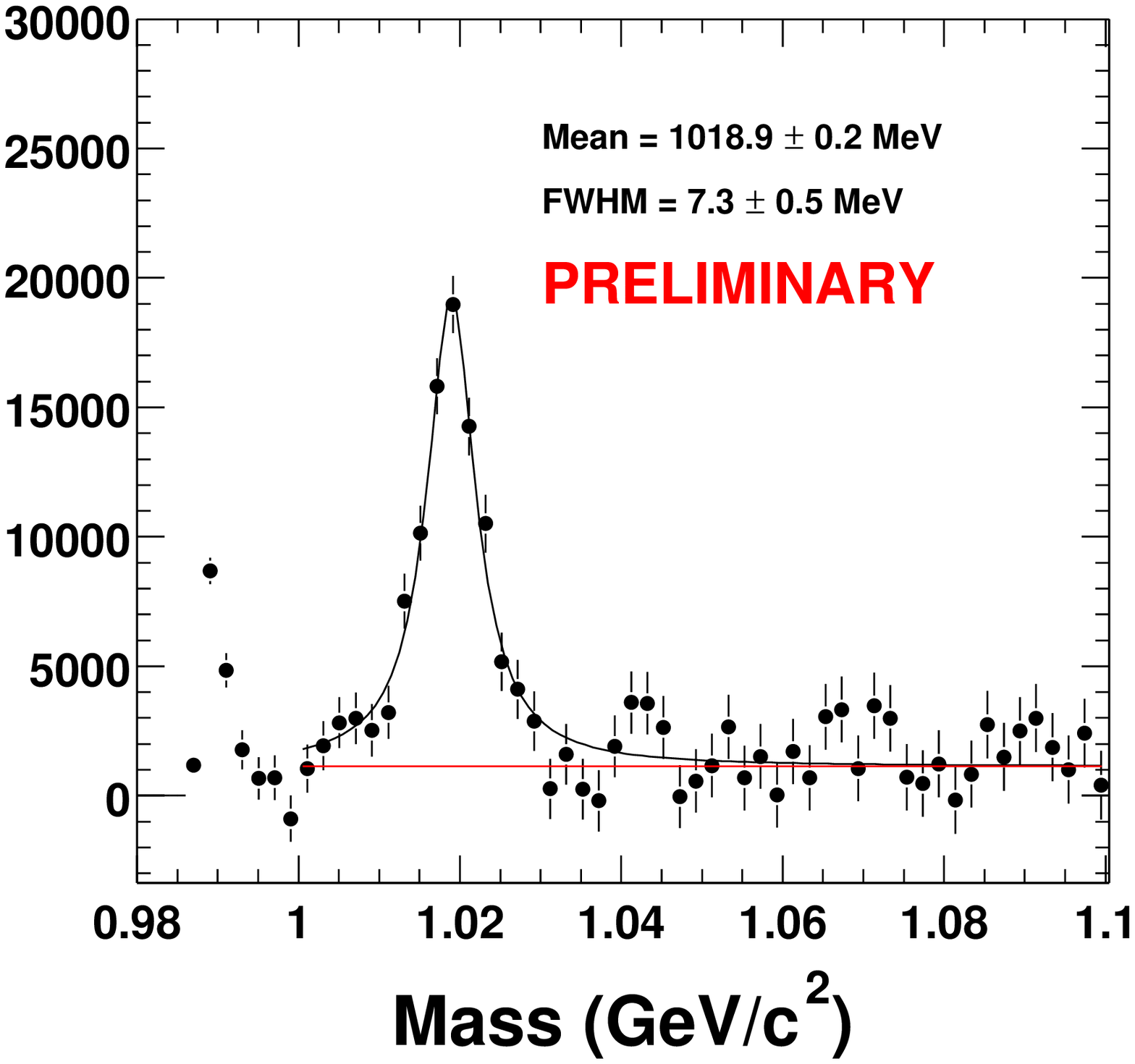,width=.5\textwidth}
\psfig{figure=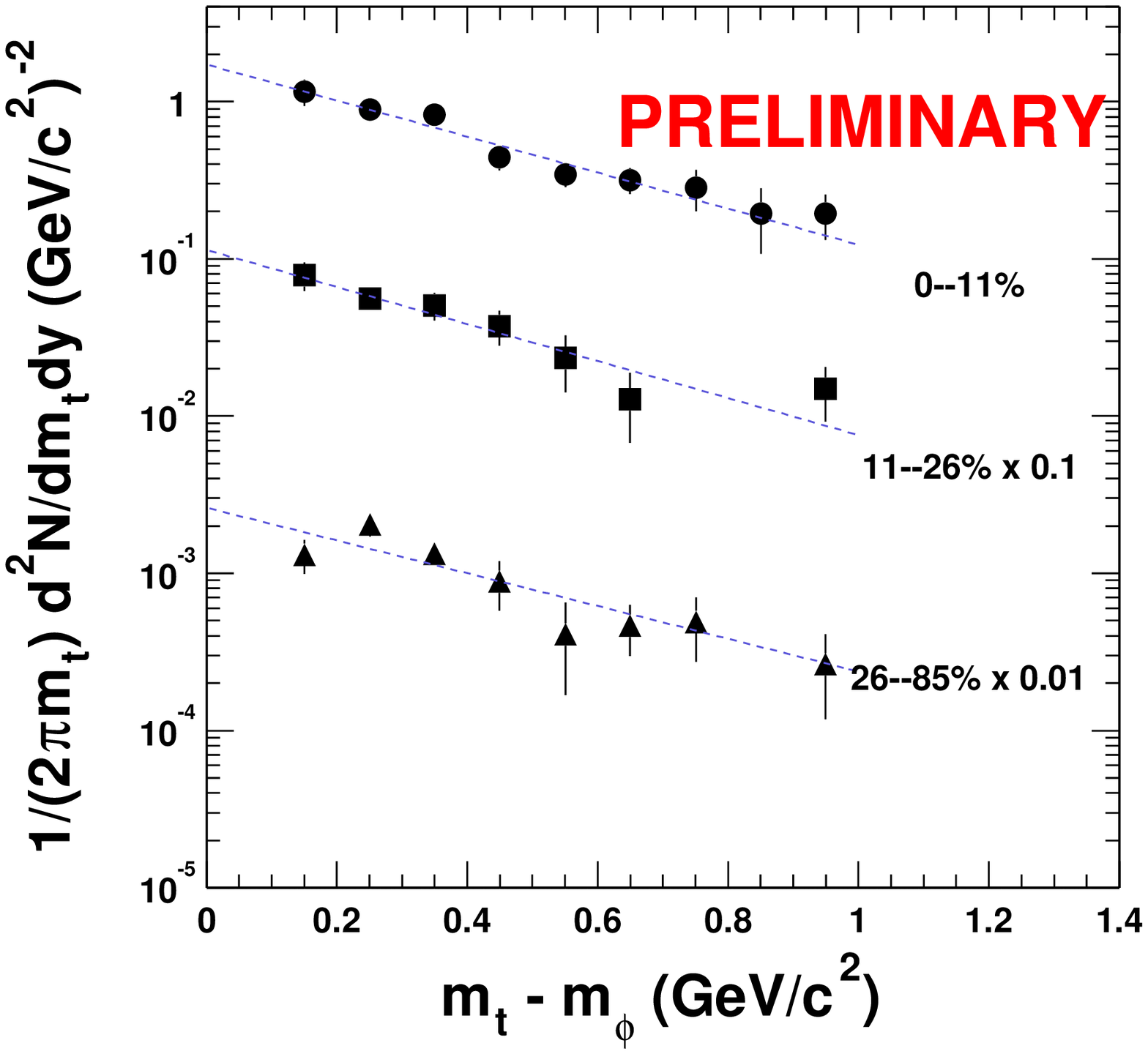,width=.5\textwidth}}
\caption{
\textbf{Left:} Invariant mass distribution for candidate $K^+K^-$ pairs
after background subtraction for the 11\%
most central collisions. The width of the invariant mass mass distribution is consistent
with the natural width of the $\phi$ convoluted with the resolution of the
TPC.
\textbf{Right:} Transverse mass distributions of $\phi$ from $Au+Au$ collision at 
$\sqrt{s_{_{NN}}}=130$ GeV for three centrality bins. 
Dashed lines are exponential fits to the data.
For clarity, data points from the 11--26\% and 26--85\% centrality bins
are scaled by 0.1 and 0.01, respectively. Error bars shown are statistical only.
}
\label{dndmt}
\end{figure}

To obtain the $\phi$ spectra, same event and mixed event distributions
were accumulated and background subtraction was done in each $(m_t,y)$ bin. 
The mixed event background $m_{inv}$ distribution was normalized to the same event $m_{inv}$
distribution in the region above the $\phi$ mass 
($1.04 < m_{inv} < 1.2\;\mathrm{GeV/c^2}$).
The raw yield in each bin was then determined by fitting the background
subtracted $m_{inv}$ distribution
to a Breit-Wigner function plus a linear background in a limited
mass range (left panel of Figure \ref{dndmt}).
A slight mismatch between the same-event and mixed-event background
distributions,
due to Coulomb interactions, track merging and other residual correlations 
leads to a structure in the
subtracted mass distribution \cite{phdthesis}.
The width of the fit to the invariant mass distribution is 
consistent with the natural width of the $\phi$ convoluted with the resolution of the TPC.


The resulting raw $\phi$ yield for each $m_t$, $y$ and centrality bin 
was then
corrected for tracking efficiency and acceptance using Monte Carlo simulations of 
physics processes and detector response. 
The reconstruction efficiency depended on
the $p_t$ of the \p, ranging from 
$\sim10$\% at $p_t = 0.46$~GeV/c  up to $\sim40$\% at $p_t = 1.4$~GeV/c (right panel of Figure \ref{de_effic}). 
The PID efficiency correction for the $\phi$ 
was calculated as the square of the single kaon PID efficiency
and included the centrality dependence of the $dE/dx$ resolution.
The corrected $\phi$ invariant yields for three centrality bins are 
shown in the right panel of Figure \ref{dndmt}. 
All results presented here are for 
reconstructed $\phi$ mesons within one unit of rapidity centered around $y = 0$ 
($|y| < 0.5$) and $0.46< p_t < 1.74$ GeV/c.
In the region where the pion band crosses the kaon band in $dE/dx$~\cite{harrisqm01}, 
corresponding to the kaon $p_t \simeq 0.8$ $\mathrm{GeV/c}$,
the signal to background ratio degrades. This leads to the larger statistical error
bars in the most central bin and prevented the extraction of the $\phi$ yields in
this region for the two lower centrality bins.
The spectra were fit to an exponential 
\begin{equation}
\frac{1}{2\pi m_t}\frac{d^2N}{dm_t dy} =\frac{dN/dy}{2\pi T(m_{\phi} + T)} e^{-(m_t - m_{\phi})/T}
\label{eq:expo}
\end{equation}
with the slope parameter $T$ and yield $dN/dy$ set as free parameters. 
The results obtained are listed in Table~\ref{tab:dndy}.
The fraction of $\phi$ mesons in the measured $p_t$ region assuming
an exponential distribution is $\sim 70\%$.
Also listed is the midrapidity ratio of the $\phi$ yield to the negative hadron ($h^-$) 
yield \cite{star_hminus} for three multiplicity bins.

The major systematic uncertainties for this analysis include contributions 
from PID efficiency and tracking efficiency. 
By varying PID and track quality requirements, we estimate a systematic  
uncertainty of $\pm$ 12\% for $T$ and $\pm$ 10\% for $dN/dy$. 
Systematic errors for the $N_{\phi}/N_{h^-}$ ratio also includes the uncertainty in the $h^-$
yield. The full range of the systematic uncertainty for $N_{\phi}/N_{h^-}$ is $\pm 20\%$.
%
\begin{table}
\centering
\begin{tabular}{lccc}
\hline
{\textbf{Event Centrality}}& {\textbf{0--11\%}} & {\textbf{11--26\%}} & {\textbf{26--85\%}} \\ 
\hline
{\slope (MeV)} & {\ct} & {\mt} & {\lt} \\
{$dN/dy$} & {\cy} & {\my} & {\ly} \\
$N_{\phi}/N_{h^-}$ & \cphipi & \mphipi & \lphipi\\
$N_{\phi}/N_{K^-}$ & \multicolumn{3}{c}{\scshape{$\Longleftarrow$ 0.10 to 0.16 $\Longrightarrow$}}\\
\hline
\multicolumn{4}{c}{\textbf{\scshape All numbers are preliminary}}\\
\hline

\end{tabular}
\caption{Mid-rapidity $\phi$ slope parameters $T$, extrapolated yield 
$dN/dy$ and the ratios $N_{\phi}/N_{h^-}$ and $N_{\phi}/N_{K^-}$ for three 
centrality bins. Both the statistical and systematic uncertainties are listed.
}
\label{tab:dndy}
\end{table}

\section{Results}
For the most central heavy ion collisions, there is an increase in 
$T$ from the AGS \cite{e917qm99} to SPS \cite{na49phi} to RHIC. 
The slope parameters, however, from $p+p$ 
collisions show no significant dependence on collision energy up to 
$\sqrt{s} = 63$ GeV \cite{na49phi,afs}.

Since the $\phi$ and anti-proton have similar masses and very different 
scattering cross sections \cite{shor}, comparison of the spectral shapes would shed light on
collision dynamics.
In the most central $Au+Au$ collisions at RHIC, 
the $\phi$ slope parameter is 
\ct(stat)\ctsu(syst) and there is
no dependence on event centrality 
 (Table~\ref{tab:dndy}) within our statistical uncertainty. 
The anti-proton slope parameter using the same fit function, however, 
measured in the $p_t$ range
$0.25 < p_t < 1$~GeV/c and without correction 
for feed-down from anti-hyperons, is found to be over 150 MeV higher 
than the $\phi$ meson slope measured in $0.5 < p_t < 1.7$ GeV/c~\cite{antiproton}.
Note that if a strong collective flow develops in the system, the measured slope
parameter should depend strongly on the fitting range.
Measurements of the $\phi$ and anti-proton over a much broader range of $p_t$ will
yield a more definitive conclusion on the dynamics of these particles.
%
\begin{figure}
\centering\mbox{
\psfig{figure=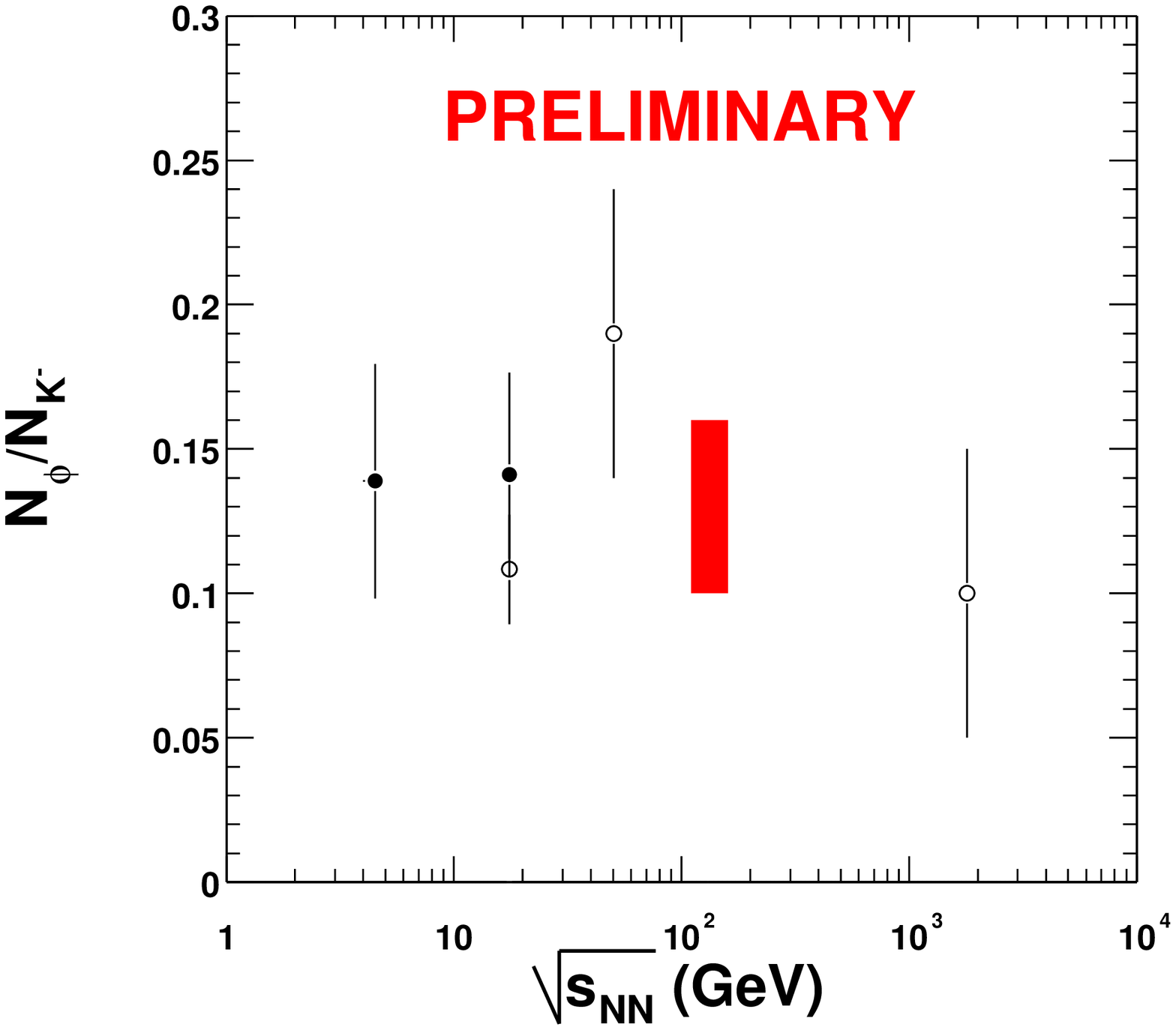,width=.49\textwidth}
\psfig{figure=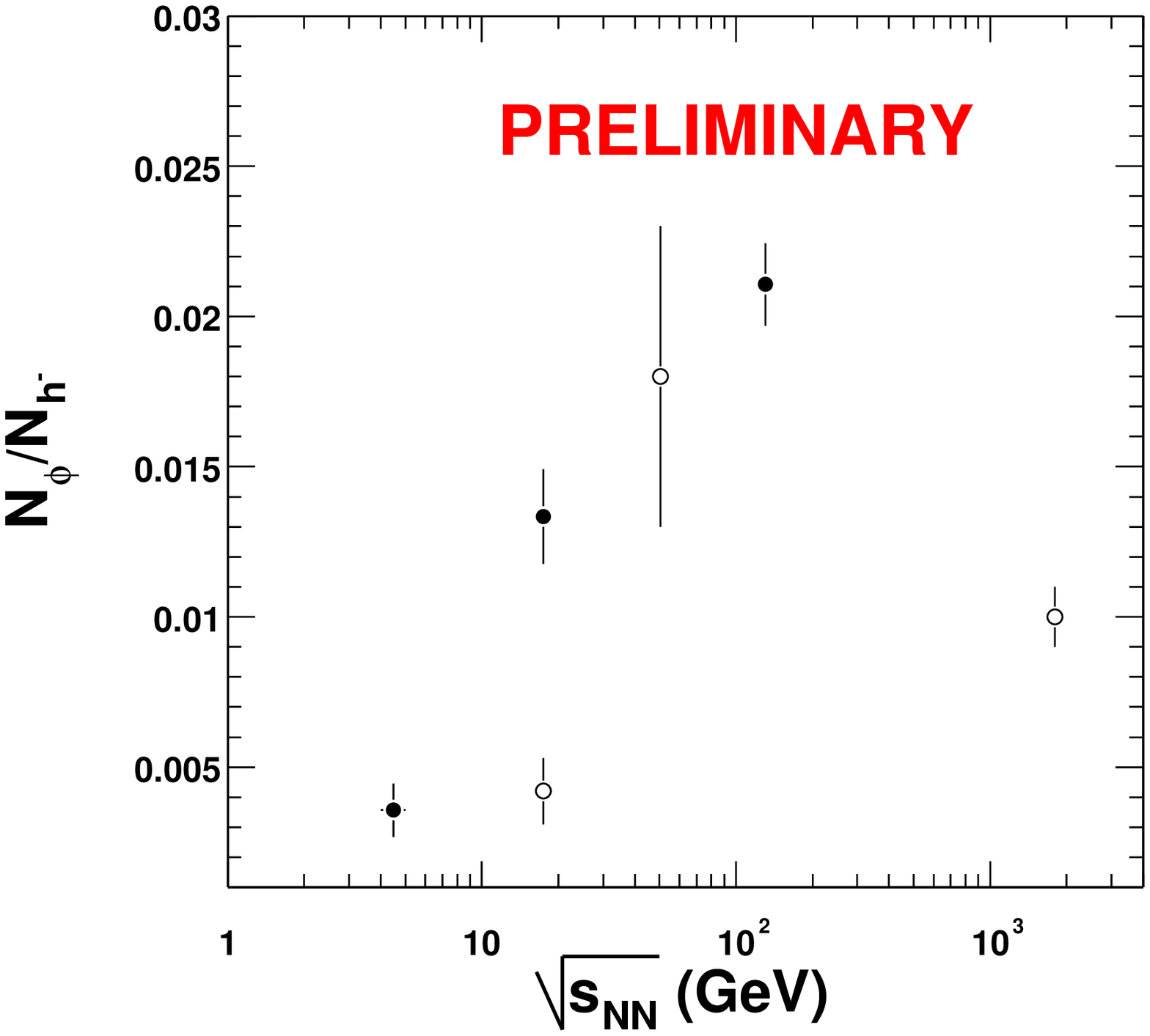,width=.49\textwidth}}
\caption{
$\sqrt{s_{_{NN}}}$ dependence of the mid-rapidity
\textbf{Left:} $N_{\phi}/N_{K^-}$ and \textbf{Right:} $N_{\phi}/N_{h^-}$ ratios. 
The data points are from the $\phi
\rightarrow K^+K^-$ decay channel. Filled symbols represent the results
extracted from the most central heavy ion collisions and the open
symbols represent the results from $p+p$ (17 and 63 GeV points) and $p+\bar{p}$ (1800 GeV) 
collisions. The shaded region in the plot of the $N_{\phi}/N_{K^-}$ ratio represents the
estimate of the ratio in $\sqrt{s_{_{NN}}} = 130$ GeV $Au+Au$ collisions at RHIC. 
Error bars shown are statistical errors only.}
\label{benergy}
\end{figure}

The energy dependence of the mid-rapidity $N_{\phi}/N_{K^-}$ ratio is shown in the 
left panel of Figure \ref{benergy}. The shaded region corresponds to our best
estimate of the $N_{\phi}/N_{K^-}$ ratio in $\sqrt{s_{_{NN}}} = 130$ GeV $Au+Au$ 
collisions at RHIC. The $N_{\phi}/N_{K^-}$
ratio has no measurable dependence on collision energy and system size
over three orders of magnitude from $Au+Au$ collisions at the AGS \cite{e917qm99}
to $p+\bar{p}$ collisions at the Tevatron \cite{tevatron_spectra}.

The energy dependence of the mid-rapidity
$N_{\phi}/N_{h^-}$ ratio is shown in the right panel of Figure \ref{benergy}.
In heavy ion collisions, $N_{\phi}/N_{h^-}$ increases 
with collision energy indicating that
$\phi$ production increases faster than $h^-$ production 
up to $\sqrt{s_{_{NN}}} = 130$ GeV. 
Although there seems to be a significant increase in $N_{\phi}/N_{h^-}$ ratio
from $p+p$ collisions between 17 and 63 GeV \cite{na49phi,afs}, the 
statistical uncertainty in the 63 GeV point is too large to determine the 
energy dependence. Note 
that the ratio at Tevatron energies ($p+\bar{p}$  at $\sqrt{s} = 1800$ GeV) 
was found to be about 0.01~\cite{tevatron_spectra}.

\section{Conclusion}
In summary, using the STAR detector we have measured 
mid-rapidity $\phi$ production 
from $Au+Au$ collisions at $\sqrt{s_{_{NN}}} = 130$~GeV. 
In the most central collisions, the $\phi$ slope parameter,
\slope$=$~\ct(stat)\ctsu(syst)~MeV, is lower than that of anti-protons in the 
measured $p_t$ region.
Within statistical uncertainty, there is no variation in $\phi$ slope parameters and
the ratio $N_{\phi}/N_{h^-}$ for the selected centrality
bins.
There is no measurable dependence of the $N_{\phi}/N_{K^-}$ on collision energy and system size
spanning three orders of magnitude.
The $\phi$ slope parameter and
the ratio $N_{\phi}/N_{h^-}$
increase from $\sqrt{s_{_{NN}}} \simeq 5$ to 130 GeV.



\begin{thebibliography}{99}
\bibitem{crosssection} The measured cross-section is approximately 90\% of 
the total inelastic Au+Au cross-section. Our centrality bins correspond to
approximately 77--23\%, 23--10\% and top 10\% of the inelastic Au+Au cross-section.
\end{thebibliography}

\end{document}